\documentclass[twocolumn,aps,prl,floatfix,showpacs]{revtex4}
\usepackage{epsfig}
\usepackage{dcolumn}
\usepackage{bm}
\begin{document}
\title{Comparative High Field Magneto-Transport of Rare Earth Oxypnictides with Maximum Transition
Temperatures}

 \author{J. Jaroszynski} \author{Scott C. Riggs}  \author{F. Hunte}  \author{A.
Gurevich} \author{D.C. Larbalestier} \author{G.S. Boebinger}
  \affiliation{National High
Magnetic Field Laboratory, Florida State University,
Tallahassee, FL, 32310}

\author{F.F. Balakirev} \author{Albert Migliori}
  \affiliation{ National High Magnetic Field Laboratory,
Los Alamos National Laboratory, Los Alamos, NM, 87545}

\author{Z.A. Ren, W. Lu,
J. Yang, X.L. Shen, X.L. Dong, Z.X. Zhao}

\affiliation{National Laboratory for Superconductivity, Institute
of Physics and Beijing National Laboratory for Condensed Matter
Physics, Chinese Academy of Sciences, P. O. Box 603, Beijing
100190, P. R. China}

 \author{R. Jin}  \author{A.S. Sefat}  \author{M.A. McGuire}  \author{B.C. Sales}  \author{D.K. Christen}  \author{D. Mandrus}
\affiliation{Materials Science Division, Oak Ridge National
Laboratory, P.O. Box 2008, Oak Ridge, TN  37831}

\date{\today}

\begin{abstract}

  We compare magneto-transport  of the  three iron-arsenide-based
   compounds ReFeAsO (Re=La,
Sm, Nd) in very high DC and pulsed magnetic fields up~to 45~T and
54~T, respectively. Each sample studied exhibits a superconducting
transition temperature near the maximum reported to date for that
particular compound. While high magnetic fields do not suppress
the superconducting state appreciably, the resistivity, Hall
coefficient and critical magnetic fields, taken together, suggest
that the phenomenology and superconducting parameters of the
oxypnictide superconductors bridges the gap between MgB$_2$ and
YBCO.

\end{abstract}


\maketitle

The recently discovered layered superconducting oxypnictides with
high transition temperatures \cite{hosono08} are based on
alternating structures of FeAs and ReO layers.  Similar to the
high-temperature superconducting cuprates, superconductivity in
oxypnictides seem to emerge upon doping of a parent
antiferromagnetic state. As the ReO planes are doped, the
ionically bonded ReO donates an electron to the covalently bonded
FeAs plane \cite{Nature}, suppressing the global
antiferromagnetism and resulting in superconductivity. Different
rare earths do, however, have an effect on the superconducting
transition temperature, $T_c$, which,  increases from a maximum of
28~K for La \cite{Nature,Sefat08} to above 40 K for Ce
\cite{Chen1} and above 50 K for Nd, Pr, Sm and Gd, respectively
\cite{Ren1,Ren2,Chen2,gadolinium}.  Unlike the cuprates, the
doping required for the onset of superconductivity, as well as the
doping at optimal $T_c$, seems to depend  on the specific rare
earth in the compound, perhaps for intrinsic reasons such as a
varying the magnetic moment or size of the rare earth atom, or a
possible role of multiple bands.

 To deduce common behaviors of the oxypnictide
superconductors, we studied three of the iron-arsenide-based
compounds ReFeAsO (Re=La, Sm, Nd) in very high magnetic fields.
All samples measured were polycrystals made by solid state
synthesis. Two were doped by partial F substitution for O (La and
Nd) and one (Sm)  by forcing an O deficiency.   The
SmFeAsO$_{0.85}$ \cite{Ren3} and NdFeAsO$_{0.94}$F$_{0.06}$
\cite{Ren1}, exhibiting a 90~\% $T_c \approx 53.5$~K and $\approx
50.5$~K, respectively, were grown  at the National Laboratory for
Superconductivity in Beijing. These samples result from
high-pressure synthesis. SmAs (or NdAs) pre-sintered powder and
Fe, Fe$_2$O$_3$, and FeF$_2$ powders were mixed together according
to the nominal stoichiometric ratio then ground thoroughly and
pressed into small pellets. The pellets were sealed in boron
nitride crucibles and sintered in a high pressure synthesis
apparatus under a pressure of 6 GPa at 1250 $^\circ$C for 2 hours.
The LaFeAsO$_{0.89}$F$_{0.11}$ sample from the Oak Ridge group has
$T_c\approx  28$~K \cite{Sefat08}, which is among the highest
transition temperatures reported for this compound at ambient
pressure \cite{Nature}.  It was grown  by a standard solid-state
synthesis method similar to that reported previously
\cite{hosono08}, from elements and binaries, with purity $> 4$~N.
Our extensive studies on morphology, connectivity, electromagnetic
granularity, phase purity, magnetic properties etc. could be find
elsewhere \cite{RefASCSmMag,yamamotoLa,EBDStak}.

Phase diagrams for fluorine-doping
\cite{hosono08,REFSmphase,REFNdphase} have been reported for each
compound: our LaFeAsO$_{0.89}$F$_{0.11}$ and SmFeAsO$_{0.85}$
samples are deemed optimally doped based upon the maximal value of
$T_c$ in the published F-doped phase diagrams, which is consistent
with the nominal F-doping level for our samples
\cite{hosono08,REFSmphase,REFNdphase}. It is important to note,
however, that the actual doping of polycrystalline samples can be
difficult to determine precisely. For example, the nominal doping
of $x=0.06$ in our NdFeAsO$_{0.94}$F$_{0.06}$ sample is
anomalously low in light of the published phase diagram, for which
an experimental value of $T_c = 50.5$~K  corresponds to the
maximum value of $T_c$, which is reported to occur at an F-doping
of around 11 \%. However, in contrast to the cuprates, the optimal
doping range for these superconductors is wide. On the other hand,
from technological point of view, it is often that the real
F-content is much smaller than the nominal for the
ambient-pressure sintered samples, since lots of F was observed to
reacts with quartz glass. However, we observe rather higher then
nominal doping, probably because our Nd sample was high pressure
sintered. This method has a better doping effect, in contrast to
ambient pressure samples which were used for the construction of
the phase diagram \cite{REFNdphase}. Despite the uncertainty in
doping levels, we believe that all three of our samples are near
optimal doping, because they exhibit transition temperatures very
near the consensus values of highest $T_c$ for these compounds at
ambient pressure.

\begin{figure}[t]
\centering
\includegraphics[width=8.5cm]{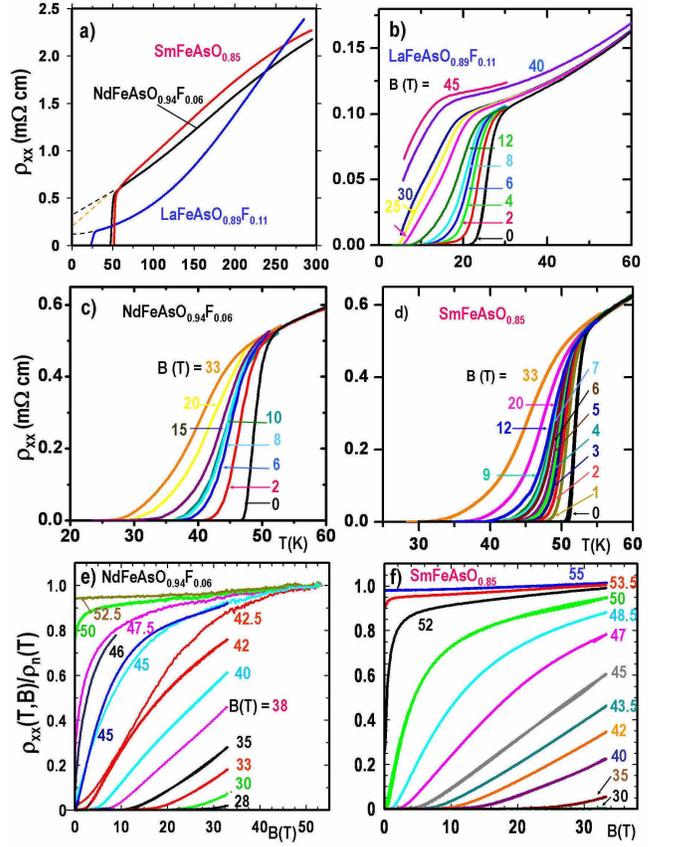}

\caption{{(a)} Longitudinal resistivity $\rho_{xx}(T)$ at $B=0$
for three members of the ReO$_{1-x}$F$_{x}$FeAs system:
LaFeAsO$_{0.89}$F$_{0.11}$ with  $T_c\simeq 28$~K,
SmFeAsO$_{0.85}$ with $T_c\simeq 53.5$~K, and
NdFeAsO$_{0.94}$F$_{0.06}$ with   $T_c\simeq 50.5$~K. Dashed lines
show normal resistance $\rho_{n}(T)$ extrapolated below $T_c$. {
(b)} $\rho_{xx}$ at various magnetic fields as a function of
temperature  for La-compound at $B = \mbox{0, 2, 4, 6, 8, 12, 20,
25, 30, 40}$ and 45~T; { (c)} Nd-compound at $B = \mbox{0, 2, 6,
8, 10, 15, 20}$, and 33~T; and { (d)} for Sm-compound at $B =
\mbox{0, 1, 2, 3, 4, 5, 6, 7, 8, 9, 12, 25}$ and 33~T. { (e)}
Normalized resistivity $\rho_{xx}(B,T)/\rho_n(T)$ of Nd sample vs.
$B$ at various $T$ measured in DC resistive magnet up to 33~T
(sweep rate 5~T/min) and  pulsed magnetic field up to 54~T. The
latter data exhibit saturation at lower temperatures which may
result from eddy current heating in this relatively big sample. {
(f)} $\rho_{xx}(B,T)/\rho_n(T)$ of Sm sample versus $B$ at various
$T$ measured in DC resistive magnet up to 33~T. }\label{fig:fig1}
\end{figure}

The longitudinal resistivity $\rho_{xx}$ and the Hall coefficient
$R_H$ in high magnetic fields were measured using a lock-in
technique in three different high-field magnets at the National
High Magnetic Field Laboratory (NHMFL): 33~T DC resistive and 45~T
hybrid magnets at Florida State University, and 54~T pulsed
magnets at Los Alamos National Laboratory.  For each experiment,
the samples were nominally rectangular prisms and the magnetic
field was applied perpendicular to the largest face of the
samples. Six electrical contacts (using either DuPont silver paint
or Epo-tek H20 E silver epoxy) were positioned around the
perimeter of the sample in a conventional Hall bar geometry

Figure~\ref{fig:fig1}~a shows $\rho_{xx}$ as a function of
temperature $T$ at $B=0$ for  the LaFeAsO$_{0.89}$F$_{0.11}$,
SmFeAsO$_{0.85}$, and NdFeAsO$_{0.94}$F$_{0.06}$ samples.
LaFeAsO$_{0.89}$F$_{0.11}$ exhibits a conventional super-linear
$T$-dependence of $\rho_{xx}(T)$.  A magnetic field suppresses
superconductivity by shifting the resistive transition to lower
$T$, reducing $T_c$ by roughly a factor of two with 30~T, as shown
in Fig.~\ref{fig:fig1}~b.

 SmFeAsO$_{0.85}$
 and NdFeAsO$_{0.80}$F$_{0.06}$
 show strikingly different behavior that is
more reminiscent of the high-temperature superconducting cuprates:
a linear temperature dependence of $\rho_{xx}(T)$, from 225~K down
to $T_c$ and a substantial broadening of the resistive transition
in a magnetic field: applying 33~T has little effect on the
high-temperature superconductivity onset, while the foot of the
transition is shifted to substantially lower temperature as shown
in Fig.~\ref{fig:fig1}~c and d.

\begin{figure}[tbh]
  \centering

\includegraphics[width=8.5cm]{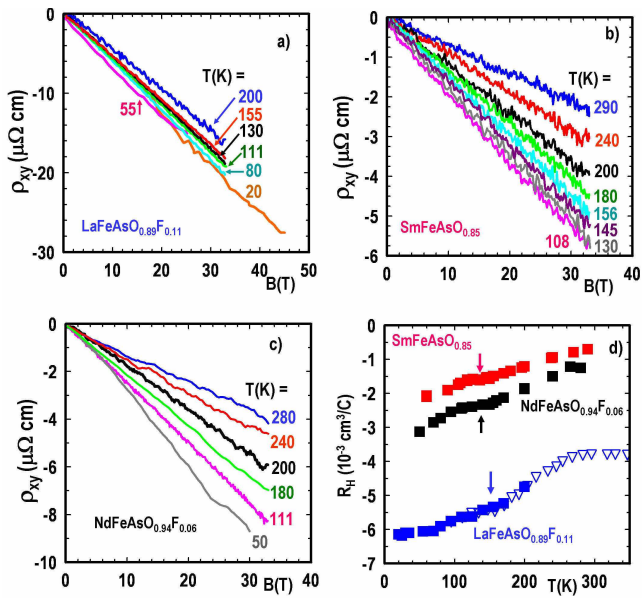}

\caption{The Hall resistivity $\rho_{xy}$ versus magnetic field
for: { (a)} La-compound, { (b)} Sm compound and { (c)} Nd-compound
at $T>T_c$. Note that all $\rho_{xy}$ show a linear dependence on
magnetic field. The low-temperature $\rho_{xy}$ for the La
compound
 are also temperature independent, consistent
with a dominant single-carrier conduction mechanism, while at
higher $T$
  all compounds show marked $\rho_{xy}(T)$
dependencies. {(d)} Hall coefficient determined from linear fits
to $\rho_{xy}(H)$ for $-33<B<33$~T or $11.5<B<45$~T (solid
squares)) and at $B=9$~T (open triangles Ref.~\cite{Sefat08}). No
systematic departure from linearity was found regardless of
temperature. Arrows show apparent inflections on $R_H(T)$
dependence around $T\approx150$~K, which may result from
structural transitions. }\label{fig:fig2}
\end{figure}

The field dependencies of the Hall resistivity $\rho_{xy}$ are
shown in Fig.~\ref{fig:fig2}~a,b, and c for  the La, Sm, and
Nd-compounds, respectively, over a wide temperature range above
$T_c$. The low temperature behavior of the La-compound at $T
< 100$~K in Fig.~\ref{fig:fig2}~a is rather conventional: $\rho_{xy}$ is linear
in $B$ and is temperature independent. If only one band
contributes to current transport,  the Hall coefficient, $R_{H}
= \rho_{xy}/B = 1/ne$, yields a carrier density $n$ of 0.07
electrons per unit formula in the La-compound at low temperatures. Here we
used reported unit cell parameters
\cite{REFunitcellsizes} with two unit formulas per unit cell. This
value  is not far from the nominal doping of $x=0.1$ per unit
formula.

At higher $T$ all three compounds exhibit a strong temperature
dependence of $\rho_{xy}$ as is evident in Figs.~\ref{fig:fig2}
a,b, and c. Note however that all traces in Fig.~\ref{fig:fig2}
retain a linear dependence on $B$ regardless of temperatures. The
observed strictly linear field dependence of $\rho_{xy}$ and no
magnetoresistance at temperatures 70-200~K indicates that
$\omega_{c}\tau << 1$, where $\omega_{c }=eB/m^*$ is the cyclotron
frequency and $\tau$ is the scattering time. At the same time the observed temperature
dependence of $R_H$ may indicate two-band effects.

A number of experimental \cite{Hunte,wenMULTI,matanoNMR} and
theoretical \cite{Ma,raghuMULTI,hanMULTI} papers have discussed
the oxypnictides as a two-band system. In this case, the primary
effect of high magnetic fields is the lateral separation of
electron-like and hole-like carriers. Often $\rho_{xy}$ for a
semimetal is not a linear function of $H$ due to the complexity of
the transport in the presence of two different carriers.  The
field dependence of $R_H$ for a two band system \cite{Kittel}
    \begin{equation}
    R_H=\frac{\sigma_1^2R_1+\sigma_2^2R_2+\sigma_1^2\sigma_2^2R_1R_2(R_1+R_2)B^2}{
    (\sigma_1+\sigma_2)^2+\sigma_1^2\sigma_2^2(R_1+R_2)^2B^2}
    \label{rh}
    \end{equation}
becomes noticeable as the parameter $\omega_c\tau=B/B_0$ becomes
of the order of unity provided that $R_1\sim R_2$ and $\sigma_1\sim \sigma_2$.
Here $B_0 = \rho_{xx}/R_H$, the indices 1
and 2 correspond to bands 1 and 2, $R_{1,2}$ and
$\sigma_{1,2}=1/\rho_{1,2}$ are intraband Hall coefficients and
conductivities, respectively, and $\rho_{xx}$ is determined by the
minimum value of $\rho_1$ and $\rho_2$ in the parallel band
connection. Taking the characteristic values of $\rho_{xx}\simeq
1$ m$\Omega$cm and R$_{H}\simeq 5\times 10^{-3}$ cm$^{3}$/C from
Figs.~\ref{fig:fig1} and \ref{fig:fig2}, we get $B_0=2000$ T,
indicating that $\omega_c\tau>>1$ only at inaccessible magnetic
fields $H
>1000$~T. The same field $B_{0}$ sets the scale for the onset of
magnetoresistance. Note that in Fig.~\ref{fig:fig1}, there is no
significant magneto-resistance above $T_c$, where the zero field
data (black line) superimpose data taken in magnetic fields. Thus,
the observed temperature dependence of
$R_H$ is consistent with
a two-band system at $\omega_{c}\tau\ll 1$ for both bands with no visible
magnetoresistance and the Hall resistivity $\rho_{xy} =B(\sigma_1^2R_1+\sigma_2^2R_2)/(\sigma_1+\sigma_2)^2$
linear in $B$.

Figure~\ref{fig:fig2}~d shows the marked temperature dependencies
of $R_{H}$, which exhibit pronounced inflections at temperatures
marked by the three arrows. These arrows indicate the temperatures
at which a structural phase transition has been reported for the
three {\it undoped} parent compounds.  Neutron data on undoped
LaFeAsO show a transition from tetragonal to monoclinic lattice
around 150~K, followed by antiferromagnetic ordering below 134~K
\cite{REFLaSPT}. SmFeAsO shows a tetragonal to orthorhombic
transition at 129~K \cite{REFSmSPT}.  For NdFeAsO, sharp jumps in
heat capacity signal structural transition at $\approx 150$~K
\cite{REFNdSPT}. Unlike the pseudogap in the cuprates, doping of
the parent compound only slightly reduces the temperature of these
anomalies in the oxypnictides as the doping approaches the range
in which superconductivity is observed
\cite{hosono08,REFSmphase,REFNdphase}. The Hall data in Fig.~2
suggest that a remnant of the structural transitions in the
undoped parent compounds manifest themselves at optimum doping as
well, however it is also possible, that this results from the
presence of undoped phase in the samples. At the same time the
overall temperature dependence of $R_H(T)$ is not understood.

\begin{figure}[tb]
\includegraphics[width=8.5cm]{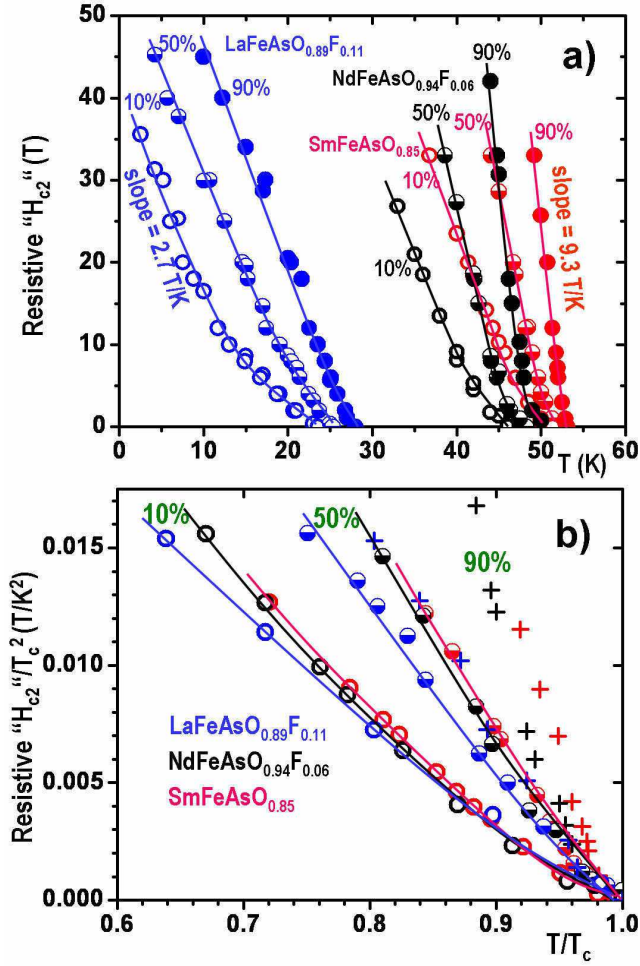}
\caption{{ (a)} Upper critical fields $H_{c2}$ versus temperature
for the Sm-, Nd- and La-compounds. with $T_c = 53.5$~K, 50.5~K,
and 28~K respectively, as determined from 90~\% transitions; $T_c
= 52$~K, 47.5~K, and 25~K for 50 \% transitions and $T_c = 51$~K,
46~K, and 23~K for 10 \% transitions. The data extracted from the
results shown in Fig.~1, show the temperatures at which the
resistance reaches 10~\%, 50~\%
 and 90~\%  of the normal state
resistance, as extrapolated linearly from the $\rho_{n}(T)$
temperature dependence above $T_c$. The solid lines guide the eye.
{ (b)} The same data divided by $T_c^2$ and plotted as a function
of reduced temperature $T/T_c$. }\label{fig:fig3}
\end{figure}

Finally, the high-field longitudinal resistivity measurements
shown in Fig. \ref{fig:fig1} enable a characterization of the
magnetic field scales required to suppress superconductivity in
the oxypnictides. Figure \ref{fig:fig3} show three characteristic
fields $H_{90}(T)$, $H_{50}(T)$ and $H_{10}(T)$ at which
$\rho_{xx}(T)$ reaches 90\%, 50\% and 10\% of the normal state
resistivity $\rho_n(T)$ extrapolated linearly from its temperature
dependence above $T_c$. For all three compounds, the data in
Fig.~\ref{fig:fig3} are reminiscent of the quasi-2D layered
cuprate superconductors \cite{Ando}. Following a previous analysis
for polycrystalline samples \cite{Hunte}, we assume that,
because of the strong angular dependence of $H_{c2}(\theta)$, the field
$H_{10}(T)$ scales like $H_{c2}^{\perp}$ parallel to the c-axis,
that is, the geometry for which the magnetic field most readily
suppresses the superconducting state. The resistivity
$\rho_{xx}(T,H)$ at the very bottom of the transition, say at
$0.5\%\rho_n$ matches the onset of the pinned critical state of
vortices below the irreversibility field $H_{irr}(T)$ extracted
from magnetization measurements \cite{RefASCSmMag} The 90\% data
in turn likely corresponds to $H_{c2}(T)$ perpendicular to the
c-axis, as only at this point has superconductivity been largely
suppressed for all orientations of the polycrystalline grains.

Shown in Fig.~\ref{fig:fig3}~b are the magnetic fields
$H_{90}(T)$, $H_{50}(T)$ and $H_{10}(T)$ normalized to the
respective values of $T_c^2$ for each compound and plotted as
functions of the reduced temperature $t=T/T_c$. All $H_{50}(T)$
and $H_{10}(T)$ data nearly collapse onto single curves, while the
$H_{90}(T)$ data do not. This behavior suggests the following
qualitative interpretation. The upper critical fields
$H_{c2}^{\perp}\sim\phi_0/2\pi\xi_a^2$ along the c-axis and
perpendicular to the c-axis, $H_{c2}^\|\sim\phi_0/2\pi\xi_a\xi_c$
are defined by the respective coherence lengths, $\xi_a$ and
$\xi_c\simeq \xi_a\Gamma^{-1/2}$, where $\Gamma=\rho_c/\rho_{ab}$
is the effective mass or resistivity anisotropy parameter. Given
the nanoscale $\xi_a$ and $\xi_c$ in oxypnictides \cite{Hunte}, we
can assume that these materials even in the present
polycrystalline form are likely in the clean limit, $\xi_a<\ell$,
where $\ell$ is the mean free path. In this case
$\xi_a\propto\xi_c\propto 1/T_c$, which yields $H_{c2}\propto
T_c^2$, consistent with the scaling of $H_{50}(T)$ and $H_{10}(T)$
curves defined by $H_{c2}^{\perp}$ in Fig. 4b for all three
compounds. In turn, the lack of scaling for $H_{90}(T)$ defined by
$H_{c2}^\|(T)$ in Fig.~\ref{fig:fig3}b may indicate that, in
addition to the change in $T_c$, the mass anisotropy parameter
$\Gamma$ also changes, as will be discussed below.

Our high-field data enable us to make several further conclusions
regarding trends in superconducting oxypnictides. From the
measured $R_H = 1/ne$ and the London penetration depth
$\lambda_0^2 = m^*/\mu_{0}ne^{2}$ at $T = 0$, we estimate the
effective mass of the carriers $m^* = \mu_0e\lambda_0^{2}/R_{H}$,
where e is the electron charge. Taking $R_H = 6.2\times10^{-3}$
cm$^{3}$/C at $T_c$ from Figs.~\ref{fig:fig1} and \ref{fig:fig2}
and $\lambda_0 =$ 215~nm from NMR data \cite{NMR}, we obtain $m^*
= 1.6m_e $ for LaFeAs(O,F) (or a slightly higher $m^* = 2.23 m_e$
for $\lambda_0 = 254$~nm taken from $\mu$SR data
\cite{REFLaMuSR}). For SmFeAsO, we take $R_H = 2\times10^{-9}$
m$^{3}$/C from Figure 2 and $\lambda_0 = 184$ nm from $\mu$SR data
\cite{REFSmMuSR}, which yields $m^* = 3.7m_e $. The increase of m*
as $T_c $ increases from 28~K for
 LaFeAs(O,F) to 53.5~K for SmFeAsO may reflect the effective
mass renormalization by strong coupling effects
\cite{REFMassRenorm}. The above one band estimate can also be applied
qualitatively to two-band systems, in which either the ratios $m_1/n_1$ and $m_2/n_2$ are not
too different, or $\sigma_1\gg \sigma_2$. In the latter case
$m^*$ corresponds to band 1, which effectively short-circuits band 2.  In turn,
the ratio $\sigma_1/\sigma_2$ can be controlled by disorder, as
has been shown for MgB$_2$ \cite{ag}.

The relative effects of vortex fluctuations can be inferred from the data
of Fig.~ \ref{fig:fig1}. As mentioned previously, the $R(T)$ curve
for  LaFeAs(O,F) mostly shifts to lower temperatures without much
change in shape upon increasing $H$. This behavior is
characteristic of the resistive transition in low-T$_{c}$
superconductors with weak thermal fluctuations of vortices. By
contrast, the $R(T)$ curves for the higher-T$_{c}$ oxypnictides in
Figs.~\ref{fig:fig1}~c and d broaden significantly as $H$
increases, similar to the behavior of the cuprates. This indicates
that thermal fluctuations of vortices in SmFeAsO and
 NdFeAs(O,F) appear to be much stronger than in  LaFeAs(O,F). The
effect of thermal fluctuations is quantified by the Ginzburg
parameter, $Gi = (2\pi k_BT_c\mu_0\lambda_0^2/\phi_0^2\xi_c)^2/2$,
$\xi_c=\xi_a\Gamma^{-1/2}$ is the coherence length along the
c-axis, $\Gamma=m_c/m_a$ is the mass anisotropy parameter in a
uniaxial crystal \cite{Blatter}.

For  LaFeAs(O,F), $\xi_c$ can be estimated from the zero
temperature $H_{c2}^{\parallel}(0) = \phi_{0}/2\pi\xi_a\xi_c$,
which yields $\xi_{c} = [\phi_{0}/2\pi
H_{c2}^\|(0)\sqrt{\Gamma}]^{1/2} = 1.2$~nm for $\Gamma = 15$
\cite{singh} and $H_{c2}^{\parallel}(0) = 60$~T \cite{Hunte}. In
this case $Gi = 3.4\times10^{-4}$ is only 50\% higher than $Gi =
2.1 \times10^{-4 }$ of clean MgB$_{2}$ (T$_{c} = 40$~K, $\xi_{a
}= 5$~nm, $\Gamma = 36$, and $\lambda_{a}/\xi_{a }= 25$
\cite{Canfield}), but some 30 times smaller than $Gi$ for the
least anisotropic cuprate, YBa$_{2}$Cu$_{3}$O$_{7-x}$
\cite{Blatter}. Since $\xi$ is inversely proportional to $T_c$,
the slope $dH_{c2}^{\parallel}/dT \propto T_{c}\Gamma^{1/2}$ near
T$_c$ in the clean limit increases as $T_c$ and $\Gamma$ increase.
The values of $dH_{c2}^{\parallel}/dT$ estimated from the slopes
of $H_{90}(T)$ in Fig.~\ref{fig:fig3} increase from 2.7 T/K for
 LaFeAs(O,F) to 9.3 T/K for SmFeAsO, indicating that
SmFeAsO is more anisotropic with $\Gamma \sim
15(9.3T_{c}^{La}/2.7T_{c}^{Sm})^{2}\simeq 65$, which exceeds the
anisotropy parameter $\Gamma\sim 25-50$ of YBCO. This estimate is
very close to $\Gamma\approx 64$ at $T_c$ inferred from the recent
torque magnetic measurements on SmFAs(O,F) single crystals, which
also revealed a temperature dependence of $\Gamma(T)$ indicative
of two-band superconductivity \cite{torque}. The impact of high
anisotropy on the Ginzburg parameter for SmFeAsO is large:
$Gi^{Sm}/Gi^{La}=
(T_{c}^{Sm}\lambda_{0}^{Sm}/T_{c}^{La}\lambda_{0}^{La})^{4}(\Gamma_{Sm}/\Gamma_{La})\simeq
35$ and $Gi^{Sm}$ is thus of the same order as $Gi$ for YBCO. Our
conclusions are also consistent with the recent measurements of
$H_{c2}$ on NdFeAs(O,F) single crystals, for which
$dH_{c2}^\|/dT=9$ T/K, $dH_{c2}^{\perp}/dT=1.85$ T/K, $\xi_a\simeq
1.85$ nm, $\xi_c\simeq 0.38$ nm, and $\Gamma\simeq 20-40$
depending on the sample purity~\cite {ndsc}. These values of $\xi$
are close to those for YBCO. Moreover, taking $\lambda_0=200$ nm,
$\xi_c=0.38$ nm and $T_c=49$ K, we again obtain $Gi\simeq
10^{-2}$, a typical Ginzburg number for YBCO. For two-band
superconductors, the above estimates of $Gi$ remain qualitatively
the same if $\Gamma$ and $\xi$ are taken for the band with the
minimum effective mass or maximum electron mobility \cite{ag}.

The relatively small value of $Gi$ in the La compound indicates
that the melting field of the vortex lattice, $H_m(T)$ is not too
different from $H_{c2}$ \cite{Hunte}. However, for Sm and
Nd-compounds, the difference between $H_{m}$ and $H_{c2}^{\perp}$
becomes more pronounced because of the much larger Ginzburg
numbers. In this case $H_{m} << H_{c2}$, even for a moderately anisotropic
superconductor, for which
    \begin{equation}
    H_m=aH_{c2}(0)\frac{T_c^2}{T^2}\left(1-\frac{T}{T_c}\right)^2.
    \label{hm}
    \end{equation}
Here $a = \pi^2c_L^4/Gi$, and $c_{L} \approx 0.17$ is
the Lindemann number for the vortex lattice \cite{Blatter}. Taking
the above-estimated value of $Gi \sim 1.3\times 10^{-2}$ for
SmFeAsO, we obtain $a\sim 0.63$. Here the melting field $H_{m}$(T)
exhibits an upward curvature near $T_{c}$ where $H_{m}$(T) is
significantly smaller than $H_{c2}(T) \approx H_{c2}(0)(1-t)$. At
lower temperatures $H_{m}(T)$ crosses over with $H_{c2}(T)$  in
the same way as in cuprates \cite{Blatter}.

In conclusion, we have measured magneto-transport in three of the
rare earth oxypnictide superconductors:
LaFeAsO$_{0.89}$F$_{0.11}$, SmFeAsO$_{0.85}$, and
NdFeAsO$_{0.94}$F$_{0.06}$. From resistivity, Hall coefficient and
upper critical magnetic fields, we conclude that
LaFeAsO$_{0.89}$F$_{0.11}$ behaves as an intermediate-$T_{c}$
superconductor like MgB$_{2}$ in which thermal fluctuations of
vortices do not significantly affect the $H$--$T$ diagram to the
extent that they do in the layered cuprates. However, the
situation is different for the higher $T_c$ oxypnictides, SmFeAsO
and  NdFeAs(O,F), which exhibit a larger mass anisotropy, enhanced
thermal fluctuations, and for which the Ginzburg parameter becomes
comparable to that of YBCO.  Thus, the series of oxypnictide
superconductors bridges a conceptual gap between conventional
superconductors and the high-temperature cuprates. As such, they
hold particular promise for understanding the many
still-unexplained behaviors of the high-$T_c$ cuprates.

The work at NHMFL was supported by the NSF Cooperative Agreement
No. DMR-0084173, by the State of Florida,  by the DOE, by the
NHMFL IHRP program (FH), and by AFOSR grant FA9550-06-1-0474 (AG
and DCL). Work at ORNL was supported by the Division of Materials
Science and Engineering, Office of Basic Energy Sciences.


\end{document}